\documentclass{pasj00}

\def\commenta{$^*$}
\def\commentb{$^\dagger$}
\def\commentc{$^\ddagger$}

\def\PublisherReidel{Dordrecht: D. Reidel Publishing Company}

\begin{document}
\SetRunningHead{T. Kato, D. Nogami, and S. Masuda}{Phase Reversal Superhumps in ER Ursae Majoris}

\Received{}
\Accepted{}

\title{Unusual Phase Reversal of Superhumps in ER Ursae Majoris}

\author{Taichi \textsc{Kato}}
\affil{Department of Astronomy, Kyoto University,
       Sakyo-ku, Kyoto 606-8502}
\email{tkato@kusastro.kyoto-u.ac.jp}

\author{Daisaku \textsc{Nogami}}
\affil{Hida Observatory, Kyoto University, Kamitakara, Gifu 506-1314}
\email{nogami@kwasan.kyoto-u.ac.jp}

\email{\rm{and}}

\author{Seiji \textsc{Masuda}}
\affil{Okayama Astrophysical Observatory, National Astronomical
       Observatory, Okayama 719-0232}


\maketitle

\begin{abstract}
   We studied the evolution of superhumps in the peculiar SU UMa-type dwarf
nova, ER UMa.  Contrary to the canonical picture of the SU UMa-type
superhump phenomena, the superhumps of ER UMa show an unexpected phase
reversal during the very early stage ($\sim$5 d after the superoutburst
maximum).  We interpret that a sudden switch to so-called late superhumps
occurs during the very early stage of a superoutburst.  What had been
believed to be (ordinary) superhumps during the superoutburst plateau of
ER UMa were actually late superhumps.  The implication of this discovery
is briefly discussed.
\end{abstract}

\section{Introduction}

   ER UMa stars (\cite{kat95eruma}; \cite{rob95eruma}; \cite{mis95PGCV};
\cite{nog95rzlmi}) are a small, but a very unusual, subclass of SU UMa-type
dwarf novae (cf. \cite{osa96review}; \cite{war95suuma}).
What most distinguishes ER UMa stars from other SU UMa-type dwarf novae
(hereafter we call them ordinary SU UMa stars) is the shortness (19--50 d)
of their supercycles (the interval between successive superoutbursts).
This gap between ER UMa stars and ordinary SU UMa stars has not been yet
filled even by recent
observations.  From the theoretical standpoint, the outburst properties
of ER UMa stars require unusually high mass-transfer rates within the
framework of the disk-instability theory (\cite{osa95eruma};
\cite{osa95rzlmi}), which is hard to achieve within the standard
framework of the evolution of compact binaries
(e.g. \cite{rap82CVevolution}).

   \citet{pat95v1159ori} was one of the first authors who questioned the
distinction between ER UMa stars and ordinary SU UMa stars.
\citet{pat95v1159ori} described that the evolution of superhumps in
V1159 Ori, one of the ER UMa stars, has the same properties as in
ordinary SU UMa stars.  Here we present previously unnoticed, totally
unexpected, time-evolution of the superhumps in ER UMa.  Similar
time-evolution of the superhumps has not been recorded in any ordinary
SU UMa stars.

\begin{table}
\caption{Times of superhump maxima.}\label{tab:shmax}
\begin{center}
\begin{tabular}{cccc}
\hline\hline
$E$\commenta & BJD$-$2400000 & $O-C_1$\commentb & $O-C_2$\commentc \\
\hline
  0 & 49744.2525 & -0.0019 & -0.0003 \\
  1 & 49744.3172 & -0.0030 & -0.0011 \\
 14 & 49745.1696 & -0.0054 & -0.0012 \\
 15 & 49745.2329 & -0.0079 & -0.0035 \\
 16 & 49745.3010 & -0.0055 & -0.0010 \\
 17 & 49745.3655 & -0.0068 & -0.0020 \\
 28 & 49746.0869 & -0.0087 & -0.0020 \\
 29 & 49746.1548 & -0.0065 &  0.0004 \\
 31 & 49746.2842 & -0.0086 & -0.0014 \\
 32 & 49746.3486 & -0.0100 & -0.0026 \\
 43 & 49747.0702 & -0.0117 & -0.0023 \\
 58 & 49748.0943 &  0.0261 &  0.0054 \\
 61 & 49748.2878 &  0.0223 &  0.0022 \\
 62 & 49748.3533 &  0.0221 &  0.0021 \\
120 & 49752.1618 &  0.0168 &  0.0072 \\
121 & 49752.2277 &  0.0169 &  0.0076 \\
167 & 49755.2412 &  0.0057 &  0.0046 \\
168 & 49755.3067 &  0.0054 &  0.0045 \\
183 & 49756.2826 & -0.0050 & -0.0032 \\
184 & 49756.3501 & -0.0033 & -0.0013 \\
197 & 49757.2051 & -0.0031 &  0.0012 \\
198 & 49757.2647 & -0.0092 & -0.0047 \\
229 & 49759.2936 & -0.0187 & -0.0087 \\
\hline
 \multicolumn{4}{l}{\commenta Cycle count since BJD 49744.2525.} \\
 \multicolumn{4}{l}{\commentb $O-C$ calculated against equation
                    \ref{equ:reg1}.} \\
 \multicolumn{4}{l}{\commentc $O-C$ calculated against equation
                    \ref{equ:reg2}.} \\
\end{tabular}
\end{center}
\end{table}

\section{Observation and Analysis}

   The observations were performed between 1995 January 26 (the next night
of the superoutburst maximum) and February 10,
using a CCD camera (Thomson TH~7882, 576 $\times$ 384 pixels, on-chip
2 $\times$ 2 binning adopted) attached to the Cassegrain focus of the 60 cm
reflector (focal length = 4.8 m) at Ouda Station, Kyoto University
\citep{Ouda}.  An interference filter was used which had been designed to
reproduce the Johnson $V$ band.  The frames were analyzed as in the
same manner described in \citet{kat95eruma} and \citet{kat96erumaSH}.
The differential magnitudes were measured against GSC 3439.1211, which
was commonly used in \citet{kat95eruma} and \citet{kat96erumaSH}.

   We first removed the linear decline trend (superoutburst plateau)
from the observed magnitudes, and removed small nightly deviations from
the linear decline by subtracting constants from nightly observations.
Barycentric corrections to the observed times were applied before the
following analysis.

\section{Timing Analysis of Superhumps}

\begin{figure}
  \begin{center}
    \FigureFile(80mm,50mm){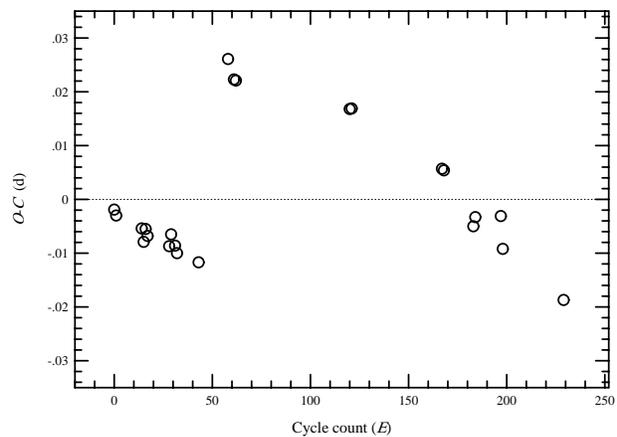}
  \end{center}
  \caption{$O-C$'s of hump maxima of ER UMa during the 1995 January--February
  superoutburst, assuming that all humps are ordinary superhumps.
  There is a striking phase jump between $E$ = 43 and $E$ = 58.
  }
  \label{fig:oc1}
\end{figure}

   We determined the maximum times of the prominent maxima from a light
curve by eye.  The averaged times of a few points close to the maximum were
used as representatives of the maximum times.  The errors of the maximum
times were usually less than $\sim$ 0.003 d.  We did not use a
cross-correlation method to obtain individual maxima because of the
variable superhump profiles.  The resultant superhump maxima are
given in table \ref{tab:shmax}.  The values are given to 0.0001 d in order
to avoid any loss of significant digits in a later analysis.
The cycle count ($E$) was first determined using the previously adopted
superhump period ($P_{\rm SH}$ = 0.06566 d).  The $O-C$'s ($O-C_1$ in table
\ref{tab:shmax}) were determined against the following linear fit to all
the maxima.  Figure \ref{fig:oc1} clearly shows that there is a striking
$O-C$ (corresponding to $\sim$0.5 phase) jump between $E$ = 43 and $E$ = 58.
This complete phase reversal clearly indicates that the humps before
$E$ = 43 and those after $E$ = 58 are essentially different in nature.

\begin{equation}
{\rm BJD (max)} = 2449744.2525 + 0.065755 E. \label{equ:reg1}
\end{equation}

\begin{figure}
  \begin{center}
    \FigureFile(80mm,50mm){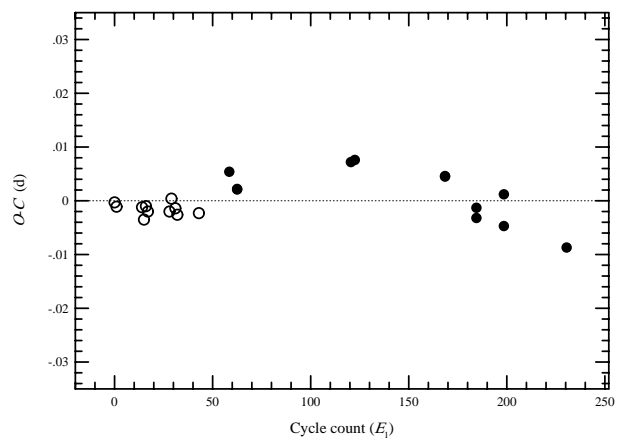}
  \end{center}
  \caption{$O-C$'s of hump maxima of ER UMa during the 1995 January--February
  superoutburst, assuming that humps with $E\geq$58 are late superhumps.
  Open and filled circles represent ordinary superhumps and late
  superhumps, respectively.
  The scales are the same as in figure \ref{fig:oc1}.
  }
  \label{fig:oc2}
\end{figure}

   As is well known, there is a superhump-type phenomenon showing
a $\sim$0.5 phase jump during the very late stage of, or shortly after
a superoutburst.  This phenomenon is called late superhumps
(\cite{hae79lateSH}; \cite{vog83lateSH}; \cite{vanderwoe88lateSH};
\cite{hes92lateSH}).  By allowing a 0.5 phase jump (or phase reversal)
between $E$ = 43 and $E$ = 58, the $O-C$ variation ($O-C_2$ in table
\ref{tab:shmax}) becomes continuous (figure \ref{fig:oc2}).  The linear
fit is represented by the following formula.

\begin{equation}
{\rm BJD (max)} = 2449744.2528 + 0.065575 E_1, \label{equ:reg2}
\end{equation}

where $E_1 = E$ for $E\leq$43 and $E_1 = E + 0.5$ for $E\geq$58.

   These results indicate that the humps with $E\geq$58 can be best
interpreted as late superhumps (we phenomenologically use this terminology
purely based on the phase jump).  In contrast to the usual evolution of
superhumps in SU UMa-type dwarf novae (\cite{vog80suumastars};
\cite{war85suuma}), ordinary superhumps in ER UMa last only for a short
time just following the superoutburst maximum and late superhumps
predominate during the most period of the superoutburst plateau.

\begin{figure}
  \begin{center}
    \FigureFile(80mm,50mm){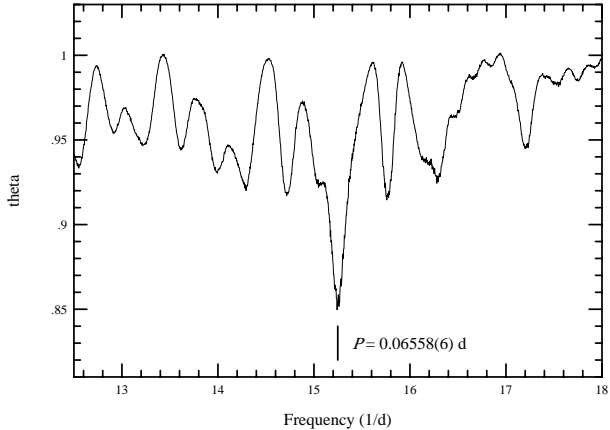}
  \end{center}
  \caption{PDM period analysis of the ordinary superhumps (i.e. BJD between
  2449744 and 2449747.5).  The period of 0.06558(6) d corresponds to the
  superhump period.
  }
  \label{fig:pdm}
\end{figure}

\begin{figure}
  \begin{center}
    \FigureFile(80mm,120mm){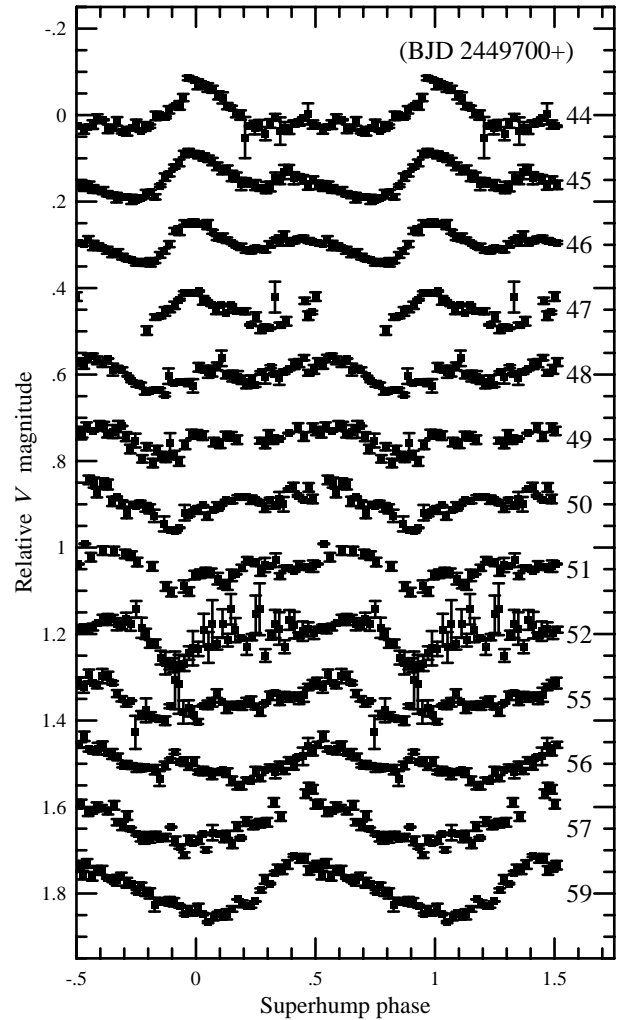}
  \end{center}
  \caption{Nightly averaged hump profiles, folded by $P_{\rm SH}$ = 0.06556 d.
  BJD 2449744 corresponds $\Delta t$ = 1 d after the maximum of the
  superoutburst.
  Ordinary superhumps (around phase$\sim$0) only appeared on the first
  four nights.  After then, late superhump (phase around $\sim$0.5)
  appeared.  A gradual shift in the maximum phase of the late superhumps
  was caused by a slow period change.
  }
  \label{fig:fold}
\end{figure}

   In order to show this transition more clearly, we first determined
the {\it true} superhump period using the maximum times with $E\leq$43.
The observed maximum times can be well expressed by a linear
ephemeris (equation \ref{equ:sh}).  A parabolic fit only yielded a
negligible quadratic term of 1.4$\pm$4.0 10$^{-6}$ d cycle$^{-1}$
(for a comparison, a fit to $E\geq$58 yields $-$2.3$\pm$0.5 10$^{-6}$
d cycle$^{-1}$).
A PDM analysis \citep{PDM} of the corresponding light curve yielded
a period of 0.065582(56) d (figure \ref{fig:pdm}).  We thus adopted
$P_{\rm SH}$ = 0.06556(2) d.  Figure \ref{fig:fold} shows nightly averaged
hump profiles folded by this $P_{\rm SH}$.
Ordinary superhumps (around phase$\sim$0) prominently appeared only
on the first four nights (until $\Delta t$ = 5 d since the start of the
superoutburst).  After then, late superhumps (phase around $\sim$0.5)
appeared, and the late superhumps were the predominant signal during
the most part of the superoutburst plateau.

\begin{equation}
{\rm BJD (max)} = 2449744.2517(3) + 0.065552(25) E. \label{equ:sh}
\end{equation}

\section{Discussion}

\begin{figure}
  \begin{center}
    \FigureFile(80mm,100mm){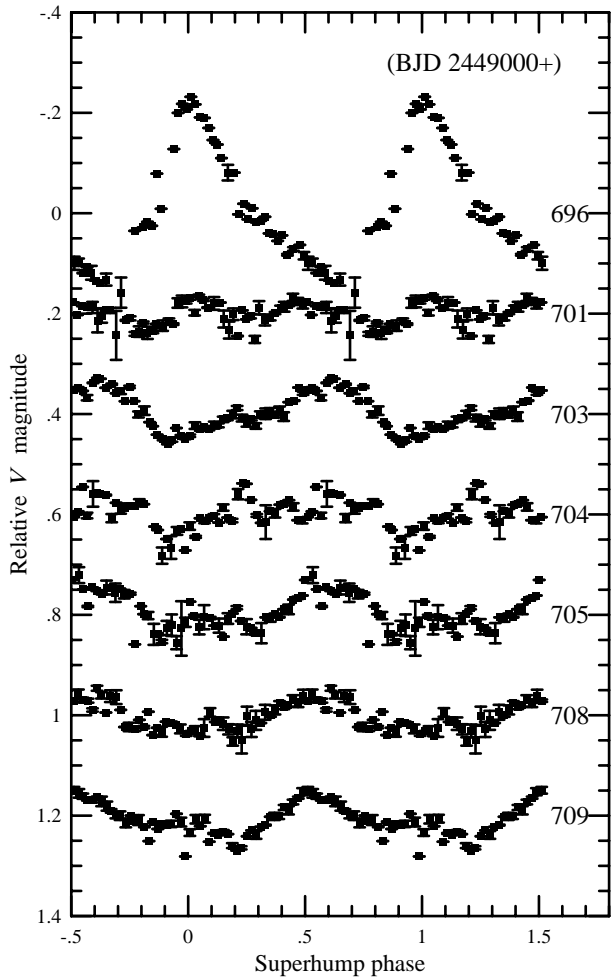}
  \end{center}
  \caption{Nightly averaged hump profiles during the 1994 December
  superoutburst.  BJD 2449696 corresponds to the maximum ($\Delta t$ = 0 d)
  of the superoutburst.  After $\Delta t$ = 5 d, the ordinary superhumps
  disappeared and late superhumps appeared.  The time-evolution of the hump
  profiles followed the same course as in the 1995 January--February
  superoutburst.
  }
  \label{fig:fold2}
\end{figure}

   Since such an early interchange between ordinary superhumps and late
superhumps is quite unexpected in any known SU UMa-type dwarf novae,
we first inspected the time-evolution of the superhumps during other
superoutbursts of ER UMa.  Figure \ref{fig:fold2} shows the time evolution
of the humps during the 1994 December superoutburst (the data are from
\cite{kat96erumaSH}).
The time-evolution of the hump profiles followed the same course as in
the 1995 January--February superoutburst.  Although the earliest stage
of the superoutburst was not observed, the 1994 January superoutburst
\citep{kat95eruma} followed the same course after $\Delta t$ = 7 d.
Thus, what had been believed to be (ordinary) superhumps during the
superoutburst plateau of ER UMa were actually late superhumps.  What
were observed as a rapidly decaying giant superhumps at the very early
stage of a superoutburst \citep{kat96erumaSH} were ordinary superhumps.
These independent observations confirmed that the evolution of the
superhumps seen during the 1995 January--February superoutburst
is a feature common to different superoutbursts.

   Late superhumps in ordinary SU UMa stars usually appear late in their
superoutbursts.  This is consistent with the widely believed interpretation
that late superhumps originate from a region close to the stream-impact
point (hot spot), whose luminosity periodically varies due to a varying
release of the potential energy on an eccentric disk
(\cite{vog83lateSH}).  A significant contribution to the light
variation from the hot spot requires a condition that the luminosity of
the accretion disk is comparable to that of the hot spot.  Generally
observed ``late'' appearance of late superhumps in ordinary SU UMa stars
is consistent with this picture in that the late superhumps become
predominant only when the luminosity of the accretion disk drastically
decays during the late stage of a superoutburst.  However, such a
condition is difficult to meet during the fully outbursting state,
in which the release of the potential energy in the disk is 10$\sim$100
times larger than that at the hot spot (see e.g. \cite{osa74DNmodel}).
The (phenomenological) late superhumps in ER UMa should therefore have
a different physical origin than in ordinary SU UMa stars, unless the
energy release at the hot spot is dramatically enhanced.  From a viewpoint
of the disk-instability model \citep{osa89suuma}, a long duration of
a superoutburst is maintained by a snow-plowing effect caused by the
tidal instability.  Since it is widely believed that the ordinary
superhumps are the manifestation of the increased tidal dissipation,
a co-existence of a long-lasting plateau phase and an early decay of
the ordinary superhumps looks like a contradiction.  We consider that
the increased tidal dissipation continues even after the initial
($\Delta t\leq$5 d) phase, and the location of the strongest tidal
dissipation moves to the opposite direction (observed as a phase
reversal) in the disk by an unknown mechanism.  If such a phase reversal
is a common phenomenon in a tidally induced eccentric disk, the origin
of (traditional) late superhumps would require reconsideration.

   We can alternatively interpret that the later ($\Delta t\geq$5 d)
hump signals are genuine superhumps and the earlier (0$\leq \Delta t\leq$5 d)
hump signals bear the same physical characteristics of the late superhumps
(i.e. originating from a region close to the hot spot on an already
eccentric disk).
This interpretation would require a burst-like enhancement of the hot spot
and a preexisting eccentric disk around the start of a superoutburst.
A rapid growth of the earlier hump signals during the outburst rise
\citep{kat96erumaSH} seems to be against this possibility, since
the release of the potential energy
is expected to be reduced as the result of a rapid disk expansion during
this stage \citep{osa89suuma}.  Since this interpretation would require
a condition keenly challenging the disk-instability model of ER UMa stars
\citep{osa95eruma}, we leave it an observational open question whether
a hypothetical, enhanced hot spot region during the early stage of
superoutbursts of ER UMa can explain the unique behavior of the ER UMa
superhumps.

   We conclude that the superhump evolution in ER UMa is by no means
typical for an SU UMa-type dwarf nova, in contrast to the previous
supposition.

\vskip 3mm

This work is partly supported by a grant-in-aid (13640239)
from the Japanese Ministry of Education, Culture, Sports,
Science and Technology.

\end{document}